\documentclass[manuscript]{aastex}

\usepackage{graphicx}
\usepackage{epsfig}
\usepackage{natbib}  

\usepackage{epstopdf}  

\begin{document}

\title{Measurements and Modeling of Total Solar Irradiance in X-Class Solar Flares}

\author{Christopher Samuel Moore\altaffilmark{1,2,3}, Phillip Clyde Chamberlin\altaffilmark{4}, and  Rachel Hock\altaffilmark{5}}

\altaffiltext{1}{Center for Astrophysics and Space Astronomy, University of Colorado, UCB 389, \\Boulder, CO 80309, USA}
\altaffiltext{2}{Laboratory for Atmospheric and Space Physics, University of Colorado, UCB 600,\\ Boulder, CO 80303, USA}
\altaffiltext{3}{Astrophysical and Planetary Sciences Department, University of Colorado, UCB 391,\\ Boulder, CO 80309, USA}
\altaffiltext{4}{Solar Physics Laboratory, NASA Goddard Space Flight Center, Greenbelt, MD 20771, USA}
\altaffiltext{5}{Space Vehicle Directorate, Air Force Research Laboratory, Kirtland Air Force Base,\\ NM 87117, USA}



\begin{abstract}
The Total Irradiance Monitor (TIM) from NASA's SOlar Radiation and Climate Experiment (SORCE) can detect changes in the Total Solar Irradiance (TSI) to a precision of 2 ppm, allowing observations of variations due to the largest X-Class solar flares for the first time. Presented here is 
a robust algorithm for determining the radiative output in the TIM TSI measurements, in both the impulsive and gradual phases, for the four solar flares presented in \cite{Woods2006}, as well as an additional flare measured on 2006 December 6. The radiative outputs for both phases of these five flares are then compared to the Vacuum Ultraviolet (VUV) irradiance output from the Flare Irradiance Spectral Model (FISM) in order to derive an empirical relationship between the FISM VUV model and the TIM TSI data output to estimate the TSI radiative output for eight other X-Class flares. This model provides the basis for the bolometric energy estimates for the solar flares analyzed in the \cite{Emslie2012} study.
\end{abstract}




\section{INTRODUCTION}

The Total Solar Irradiance (TSI) is the energy released by the Sun across the entire electromagnetic spectrum. TSI contributions due to solar flares are normally masked by other solar variations, such as acoustic oscillations and granulation, which occur on time scales  similar to the initial rapid bolometric enhancement from solar flares. In the last decade, large solar flares have been detected from the TSI signal directly from the Total Irradiance Monitor \citep{Kopp2005a, Kopp2011}   
 and through an epoch superposition technique using the Variability of solar IRradiance and Gravity Oscillations (VIRGO) experiment \citep{Kretzmar2010}.
 Solar flares that exhibit measurable visible light enhancements have been deemed "white light flares" \citep{Hudson1992}.
It is possible that these white light emissions contribute large amounts of energy to the overall flare energy budget. Their contribution to the TSI energy budget is still an ongoing investigation. This paper aims to help constrain the variation in the TSI due to flares.

 Solar flares are the dominant cause of short time scale Vacuum Ultraviolet (VUV, 0.1 - 200 nm) emission. VUV wavelengths normally contribute a small portion to the TSI, but can vary by a large factor during solar activity. These emissions are one of the major drivers and energy input source of the Earth's upper atmosphere. The increase in VUV irradiance and charged particles from the Sun can effect Earth's thermosphere, mesosphere and ionosphere \citep{Qian2011}.
Radio transmissions can be interrupted and Global Positioning Systems (GPS) can be disturbed. An increase in knowledge of the flare VUV and TSI energy release is desired for studies of comparing various energies in Solar Eruptive Events (SEEs), such as was done in \cite{Emslie2012}.

 Physically, a solar flare is defined as an event where stored magnetic energy is released via an instantaneous (less than a second) rearrangement of stressed magnetic field lines, presumably magnetic reconnection, that accelerates charged particles both toward the solar suface and outward into space \citep{Hudson2011}.
The radiative enhancement stems from an increase in particle flux in the solar atmosphere and can be partitioned into two phases, an impulsive and a gradual phase.  The initial flare energy release accelerates charged particles inward towards the more dense plasma by the solar surface.  This energy input to the lower solar atmosphere heats the plasma and drives an increase of the hard X-ray flux, via Bremsstrahlung. The heated particles confined within the magnetic loops subsequently radiate thermally in soft X-rays and in some portions of the VUV. 

The initial hard X-ray temporal enhancement is called the impulsive phase and the gradual phase is the radiative cooling of the heated plasma in soft X-rays and VUV. The impulsive phase generally lasts for 5 - 10 minutes and the gradual phase peaks shortly after the impulsive phase and can last for several hours. This observed relation between the hard and soft X-ray components, deemed the Neupert Effect \citep{DennisZarro1993, Veronig2002, Benz2008, Hudson2011}
originates from the work of \cite{Neupert1968}, who noticed that the integral of the centimeter radio flux roughly matched the soft X-ray flux. The acceleration of electrons outward to relativistic speeds generating radio emission is linked to the hard X-ray emission fom the inward penetrating electrons.   Investigations of white light flares in spectral lines and continuum bands with multiple satellites demonstrate their temporal and spatial coherence with hard X-ray emission \citep{Martinez2011, Watanabe2013}.
This supports the case of white light emission occurring primarily during the impulsive phase.

This study is focused on the radiative aspect of large X-Class solar flares. An X-Class flare is a flare that has a peak irradiance greater than 10$^{-4}$ W m$^{-2}$ in the 0.1 - 0.8 nm channel of the Geostationary Operational Enviromental Satelites (GOES) X-Ray Sensor (XRS) \citep{Garcia1994}. 
In this paper we present results on decomposing flare components into each of the two phases in the VUV and in the TSI for the five flares obsereved by TIM. We also use a VUV empirical model as a proxy for un detected solar flare TSI variations.  Our updated algorithm is used to revise values in the study by \cite{Emslie2012}. In Section 2 we discuss the data and empirical model used for this study. The time series fitting procedure used to fit the empirical model to the the observed TSI data is and described in Section 3. Section 4 explains how these results are used in estimating the total X-Class solar flare radiative output during TIM eclipse periods. We discuss results and future work in Sections 5 and 6 respectively.



\section{DATA}

\subsection{Total Irradiance Monitor} \label{bozomath}

Total Solar Irradiance (TSI) data was obtained from NASA's SOlar Radiation and Climate Experiment (SORCE) mission by the Total Irradiance Monitor (TIM).
TIM has been taking solar data since SORCE was launched in January of 2003 \citep{Kopp2005b}. This instrument sets the two criteria in determining which flares are to be studied for this analysis:
1) flares that have occurred since January of 2003 and 2) flares that have an observed signal response in the TSI. Table 1 lists the five flares that met this criteria. The SORCE spacecraft is in Low Earth Orbit (LEO), which takes about 90 minutes to complete. This orbit leaves TIM with roughly 45 minute windows of solar observing when not inside of Earth's shadow (non-eclipse periods). The TIM has a time cadence of 50 seconds. The SORCE TIM is the first instrument to have the ability to observe solar
flares in the TSI \citep{Kopp2005a, Kopp2011}, partly due to this time cadence. The TIM rigorous calibration and metrology of all it's individual components are designed to attain a 100 ppm (0.010\%) combined standard uncertainty in the TSI \citep{Kopp2005b}. This absolute precision gives TIM the sensitivity necessary to observe solar flares.  Each TIM data point has an uncertainty of roughly $\sigma_{TSI}$ = 2.72 mW m$^{-2}$, corresponding to 2 ppm (1 ppm = 1,360/10$^{6}$ W m$^{-2})$.

\subsection{Flare Irradiance Spectral Model} \label{bozomath}

The Flare Irradiance Spectral Model (FISM)  \citep{Chamberlin2007, Chamberlin2008}
was used to determine the solar irradiance in the VUV wavelengths from 0.1 - 190 nm. FISM is an empirical model that estimates the solar irradiance in the VUV wavelengths with 60 second temporal resolution and 1 nm spectral resolution from 0.1 - 190 nm. These qualities allow the inclusion of variations due to solar flares, solar rotation and solar cycle variations. FISM  utilizes data from the Thermosphere Ionosphere Mesosphere Energetics and Dynamics (TIMED) Solar Extreme Ultraviolet Experiment (SEE) \citep{Woods2005}
and the Upper Atmospheric Research Satellite (UARS) SOlar STellar Irradiance Composition Experiment (SOLSTICE)  \citep{Rottman1993}.
FISM also includes the derived impulsive and gradual phases for solar flares by using the irradiance and the time derivative from the GOES XRS 0.1 - 0.8 nm channel. We analyzed 13 solar flares of Solar Cycle 23 in the VUV consistent with the Neupert effect \citep{Neupert1968} (Equation 1). The version of the Neupert effect that we employ takes only the positive values of the soft X-ray flux time derivative, $F_{sxr}$, which agrees with the hard X-ray flux, $F_{hxr}$,

\begin{equation} \label{eq:eps}
\frac{d(F_{sxr})}{dt}  \simeq F_{hxr}
\end{equation}

This relation holds when the magnetic loop plasma is heated by the influx of accelerated particles and cooling by radiation and conduction are negligible \citep{Benz2008}.


\section{DATA ANALYSIS}

The total radiative energy in the VUV wavelengths of each flare observed by GOES XRS can be modeled with FISM. The FISM formulation, calculates a daily flux component separately along with the individual flux contributions from the impulsive and gradual phases. Therefore, contributions from both the impulsive and gradual phases are already known and seperate in FISM. The fluence (joule m$^{-2}$) for each phase is computed by integrating the irradiance (W m$^{-2} $) of the impulsive ($IP$) and gradual phases ($GP$) over their respective temporal durations of the flare
and integrating over the VUV wavelengths (0.1 - 190 nm) as in Equation 2. 


\begin{equation} \label{eq:eps}
\newcommand{\ud}{\mathrm{d}}
E_{VUV} = f \int_{t_{start}}^{t_{end}}\int_{\lambda_{min}}^{\lambda_{max}} I_{FISM, \,IP + GP}(\lambda,t) \, \ud \lambda  \, \ud t
\end{equation}

Where $t_{start}$ and $t_{end}$ are the start and end times of each flare phase respectively. The low and high wavlength bounds are $\lambda_{min}$ and $\lambda_{max}$ respectively. The irradiance contributions computed here are solely from the flare enhancement and not from the rest of the Sun. After the fluence is obtained, Equation 3 gives the conversion, $f$, in units of ergs m$^{2}$ joule$^{-1}$,  that must be applied to convert from irradiance units of W m$^{-2}$ at the Earth to ergs at the sun by multiplying 



\begin{eqnarray} \label{eq:eps}
f = (1 \, AU)^{2} \, 2\pi * 10^{7} 
\nonumber\\
f = 1.406 * 10 ^{30}  
\end{eqnarray}

where $AU$ is the astronomical unit in meters.
This conversion assumes a uniform angular distribution of the flare energy release. A similar calculation was performed in \cite{Woods2006}.
This algorithm gives an approximation of the energy release in the VUV wavelengths during each phase of the flare with similar accuracies to those of FISM flare estimates. 
The VUV radiative energies for all flares in this paper are given in Table 1 and Table 2.

An amplitude fitting routine employing the FISM model was developed to fit the time profile of flares observed in the TIM TSI to obtain the bolometric radiative energy output due to solar flares. The TIM TSI data was fit to the scaled impulsive and gradual phase FISM VUV curves to acquire the energy release in each phase. The sum of the energies from each phase gives the total energy from the flare in the TIM TSI. The GOES XRS 0.1 - 0.8 nm channel was used as a temporal proxy for the gradual phase and the time derivative of this channel is used as the temporal proxy for the impulsive phase. This procedure is the application of the Neupert Effect. The TIM TSI flare signal is isolated by subtracting the background TIM TSI irradiance. A linear relationship is found from the irradiance values before and after the flare profile and the resulting background fit is subtracted off, see Figure 1. To analyze the TIM TSI flare signal, the GOES XRS (from the FISM VUV), impulsive and gradual phase proxies are scaled such that their sum matches the time signature of the background subtracted TIM TSI data, see Equation 4, Figure 2 and Figure 3. This fit is accomplished by two separate algorithms, an analytical and a numerical one. The subscript, $t$, in the definitions refer to time in the following expression. The analytical algorithm solves for the scaling coefficients of the impulsive phase $I_{t}$, with coefficient $a$,  and gradual phase $G_{t}$, with coefficient $b$, of the time profiles exactly by a least squares calculation \citep{BevingtonRobinson2003, NumericalRecipies} so that the difference between the observed data and model function, $Z_{t}$ is minimized. 

\begin{equation} \label{eq:eps}
Z_{t} = aI_{t} + bG_{t}
\end{equation}

The values obtained, by mathematical definition, will be the best coefficients
if the errors in the data are gaussian distributed and the errors are relatively small compared to the data value. Our error bars (solely from instrument precision and background subtraction), as displayed in Figure 2 and Figure 3, are small compared to their corresponding values. The data value errors are not necessarily gaussian distributed, but this merit function still gives a good estimation of the best scale factors.

The values obtained by this methodology can be positive or negative, where negative scale factors for irradiance are non-physical.  A negative scale factor was encountered for the impulsive phase of the flare on 2005 September 7, due to a very 
small impulsive phase. In addition to the analytical algorithm, a numerical minimization of chi-square was conducted by the exploration of parameter space along the minimum "bowl" for a predefined grid of values for the a and b coefficients. This method is limited by the choice of gridded values for a and b that are explored. Generally, the analytic and numerical scale factors are consistent with each other. The analytic algorithm for scale factors were used for all flares in this paper, except for the 2005 September 7 flare for the aforementioned reason.  The scaled GOES time series were integrated with respect to time and the conversion factor (Equation 3) is applied to obtain the TSI contribution for each flare phase at the Sun. The radiative energies for five flares with an observable TIM TSI signature are in Table 1.



\section{RECONSTRUCTED TSI FLARE ENERGY RELEASE}

The radiative output from X-class solar flares have large contributions from the VUV wavelength emissions, $ \sim$ 41\% for the gradual phase of center flares,
 $\sim$ 23\% for the gradual phase for limb flares. We take limb flares to have an East-West location $> 70\,^{\circ}$. Using the 5 flares seen in the TIM TSI and estimated VUV emission simultaneously obtained
from TIMED SEE and GOES respectively, the TSI flare irradiances during TIM eclipse periods can be reconstructed. This method can be used to estimate the position dependent total radiative output from flares, which is described in this section. 

The average position dependent fractional energy contribution to the flare from the VUV to the total radiative output is found for each flare.
The inverse of these values are used to scale the VUV energy output for flares not observed due to TIM eclipse periods (when TIM is behind the Earth with respect to the Sun), to obtain an estimation of the total radiative output.
This model is still limited by low statistics (5 solar flares), but as more large flares occur, more robust algorithms can be developed. The impulsive phase, $A_{Imp}$, and gradual phase $B_{Grad}$, scale factors for the time integrated VUV irradiance used to approximate the time integrated TSI 
flare output are given in Equation 5 for 
the impulsive phase and Equation 6 for the gradual phase. The total number of limb or center flares is given by $n$, the index for each individual flare in each category is $i$, $VI$ is the VUV impulsive phase energy, $TI$ is the TSI impulsive phase 
energy, $VG$ the VUV gradual phase energy and $TG$ the TSI gradual phase energy. 

\begin{equation} \label{eq:eps}
A_{Imp} = \left(\frac{1}{n} \sum_{i=1}^{n} \left(\frac{VI_{i}}{TI_{i}}\right)\right)^{-1}
\end{equation}

\begin{equation} \label{eq:eps}
B_{Grad} = \left(\frac{1}{n} \sum_{i=1}^{n} \left(\frac{VG_{i}}{TG_{i}}\right)\right)^{-1}
\end{equation}

These calculations are done separately for center and limb flares. The sclae factors for center flares are $A_{Imp} = 10.3$ and $B_{Grad} = 2.4$ for the impulsive and gradual phases respectively. Limb flare scale factors obtained were $A_{Imp} = 3.8$ and $B_{Grad} = 4.3$.



\section{RESULTS AND CONCLUSION}

The TSI energy estimations are the same order of magnitude as previously published values for X-class solar flares ($ 10^{31}$ to $10^{32}$ ergs) and have accuracies consistent to the order given
by the uncertainties. TSI radiated energies using the previously described algorithm vary somewhat from the values given the TIM analysis by \cite{Woods2006} by up to a factor of 1.5, where an additional flare on 2006 December 6 is included here that occurred after publication of the aforementioned reference. This is believed to be due to a more robust
fitting routine for both the impulsive and gradual phase scaling, but more importantly to the improved background fitting routine that can easily give a factor of 2 in the final results due to the long tail in the 
decay phase that flares have. The later is most apparent in the 2005 September 7 flare that is dominated by gradual phase emission. 

Our results are also consistent within the $1 \sigma$ uncertainties of the TSI flare contributions derived in the analysis of \cite{Kretzmar2010} using the PMO and DIRAD radiometers from the Variability of solar IRradiance and Gravity
Oscillations (VIRGO) experiment on the SOlar and Heliospheric Observatory (SOHO) mission \citep{Frohlich1997}.  As anticipated, the gradual phase contains the bulk of the energy released in solar flares as compared to the impulsive phase (an average of 7.1 times larger),  due to the gradual phase lasting much longer. In Table 1 and Table 2, demonstrates variations in the radiative output of the flares observed at Earth due to center-to-limb variations of optically thick emissions \citep{Worden2001, Chamberlin2008}. Limb flares have an expected lower energy output because the XUV wavelengths from optically thin coronal emissions, dominate the energy release and are unaffected. The optically thick emissions are reduced due to the extended path length through the solar atmosphere. Flares near disk center are more intense than the limb flares in the TSI. This center-to-limb correction can also be applied in reverse to get the TSI radiated output for a limb flare as if it actually occurred on the center of the solar disk. The modeled values are consistent with values given for two flares that occurred before 2003 that were analyzed in  \cite{Emslie2004}.

 This model can be extended for use on flares lower than X-Class, which was done in an earlier version of this analysis in the investigsation by \cite{Emslie2012}. In that study, the reconstructed TSI was computed for the majority of X-class flares of Solar Cycle 23 that were not observed by TIM, due to eclipse periods or flares that occurred before the launch of TIM in 2003. This matured version provides a more accurate estimate of the radiated energy for the energetic comparisons of solar eruptive events. The revised values given here are in general larger by roughly a factor of 2 than those in  from the \cite{Emslie2012} project, due to the modified processes to subtract the highly variable background, also as equally important, the time interval in which the fit is performed and the selection of adequate start and end times for temporal integration of the TSI data. 

These newer values are consistent with the conclusions derived in the comprehensive study of \cite{Emslie2012}. The bolomertic radiated energy is still over an order of magnitude less than the stored non-potential (free) magnetic energy in the studied active regions. This still agrees with the belief that the available magnetic energy is more than sufficient to power the flare components. The bolometric radiated energy is still less than the Coronal Mass Ejection (CME) kinetic energy. Moreover, the bolometric radiated energy is comparable to the lower limits of the flare-accelerated particles, where the latter could be up to an order of magnitude greater due to the lower energy cut-off being masked by the thermal plasma radiating in soft X-rays. The conclusion persists that there is sufficient energy in accelerated particles to  power the radiative energy release. Finally, the correlation between the soft X-ray plasma and bolometric energy is preserved.



\section{DISCUSSION AND FUTURE WORK}

A full in-depth knowledge of the energy distribution of solar flares across the entire electromagnetic spectrum is difficult to obtain due to lack of simultaneous observations  by the instruments at the time of analysis. These modeled results are the best estimations currently achievable, but are limited to the current TIM TSI data and spectrally to the VUV region. We have broken down the radiative contribution of the VUV (0.1 - 190 nm) wavelengths to the total radiative energy release for the largest solar flares, partitioned them into two separate phases and have implemented an empirical model for bolometric flare energy reconstruction. In the future, more steps will be taken to pursue increased knowledge of the spectral energy distribution of flares along with impulsive and gradual phase composition.  \\



\textbf{Acknowlegments.} The authors would like to thank the Laboratory for Atmospheric angood Space Physics (LASP) for their funding and support, as well as support from the NSF REU program.



\appendix

\section{APPENDIX}
 For completeness, our analytic fitting method is described here. The subscript, $t$, in the definitions refer to time in the following expressions. The analytical algorithm solves for the scaling coefficients of the impulsive phase $I_{t}$, with coefficient a,  and gradual phase $G_{t}$, with coefficient b, of the time profiles exactly by a least squares calculation similar to the one in \cite{NumericalRecipies}. The algorithm is as follows. For a model function, $Z_{t}$, that is linear in the parameters $a$ and $b$

\begin{equation} \label{eq:eps}
Z_{t} = aI_{t} + bG_{t}
\end{equation}

the coefficients that give the optimal approximation to the actual observed TSI data, $D_{t}$, with $N$ discrete data points, and propogated uncertainties $\sigma_{t}$,
can be found by composing a chi-square merit function, differentiating this function with respect to the 
parameters $a$ and $b$ seperately, setting them equal to zero and then solving these two coupled equations for the coefficients a and b (Equation A2 -  Equation A11).

\begin{equation} \label{eq:eps}
\chi^{2}(a,b) = \sum_{t=1}^{N} \left(\frac{D_{t} - aI_{t} - bG_{t}}{\sigma_{t}^{2}}\right)^{2}  \\
\end{equation}

Partial differentiate with respect to a and b, set equal to zero for two seperate equations\\
\begin{equation} \label{eq:eps}
\frac{\partial (\chi^{2}(a,b))}{\partial a} =  \sum_{t=1}^{N} -2 \left(\frac{D_{t} - aI_{t} - bG_{t}}{\sigma_{t}^{2}}\right) I_{t} = 0  \\
\end{equation}

\begin{equation} \label{eq:eps}
\frac{\partial (\chi^{2}(a,b))}{\partial b} =  \sum_{t=1}^{N} -2 \left(\frac{D_{t} - aI_{t} - bG_{t}}{\sigma_{t}^{2}}\right) G_{t} = 0  \\
\end{equation}

we use the substitutions,

\begin{eqnarray}
S_{II} = \sum_{t=1}^{N} \left(\frac{I_{t}}{\sigma_{t}} \right)^{2} \\
S_{DI} = \sum_{t=1}^{N} \frac{D_{t}I_{t}}{\sigma_{t}^{2}} \\
S_{IG} = \sum_{t=1}^{N} \frac{I_{t}G_{t}}{\sigma_{t}^{2}} \\
S_{DG} = \sum_{t=1}^{N} \frac{D_{t}G_{t}}{\sigma_{t}^{2}} \\
S_{GG} = \sum_{t=1}^{N} \left(\frac{G_{t}}{\sigma_{t}}\right)^{2} 
\end{eqnarray}

and then solve for a and b.

\begin{equation} \label{eq:eps}
a = \frac{S_{DI}S_{GG} - S_{DG}S_{IG}}{S_{II}S_{GG} - (S_{IG})^{2}}
\end{equation}

\begin{equation} \label{eq:eps}
b = \frac{S_{II}S_{DG} - S_{DI}S_{IG}}{S_{II}S_{GG} - (S_{IG})^{2}}
\end{equation}

This gives the impulsive and gradual phase scale factors, $a$ and $b$, respectively.




\bibliographystyle{apj}





\begin{figure}      
   \centerline{\hspace*{-0.0\textwidth}
               \includegraphics[width=0.9\textwidth]{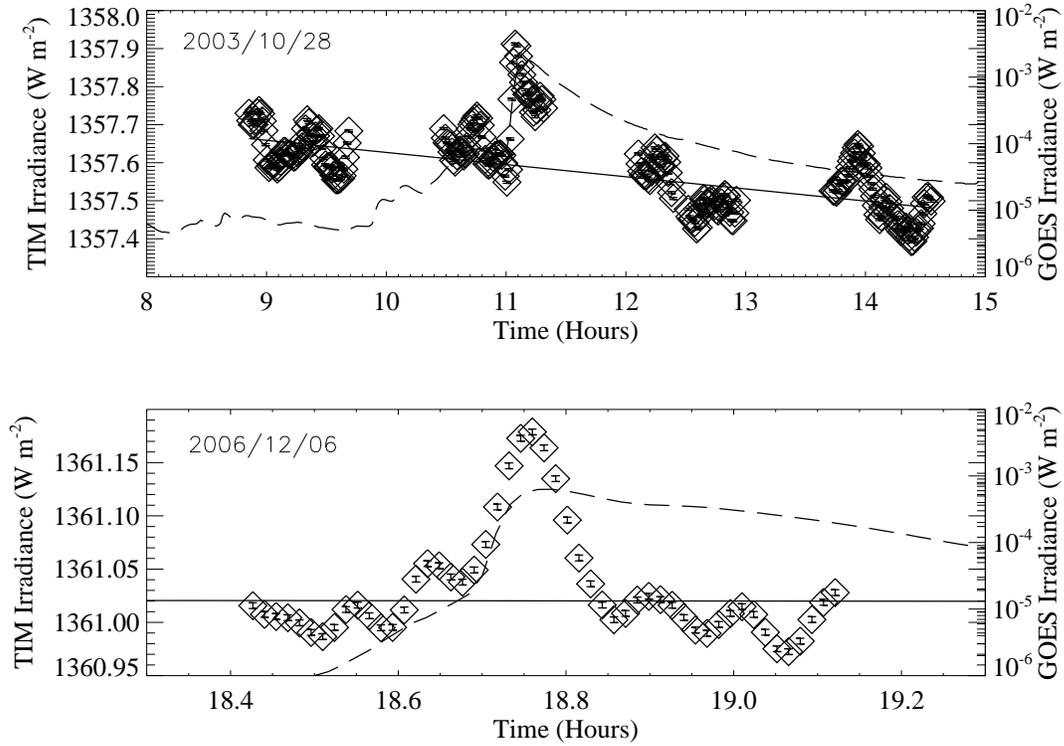}
              }
               \vspace{0.02\textwidth}   
\caption{
TSI Background subtraction for the 2003 October 28 flare and the 2006 December 6 flare that was not in \cite{Woods2006}. The solid line is the linear fit to the fluctuating background. The diamonds are the TIM TSI data with error bars from the instrument uncertanties only. The dashed line is the GOES XRS 0.1 - 0.8 nm, 1 minute time cadence data.}
\label{Oct_28_Dec_6background}
\end{figure}


\clearpage



\begin{figure}      
   \centerline{\hspace*{-0.0\textwidth}
               \includegraphics[width=0.9\textwidth]{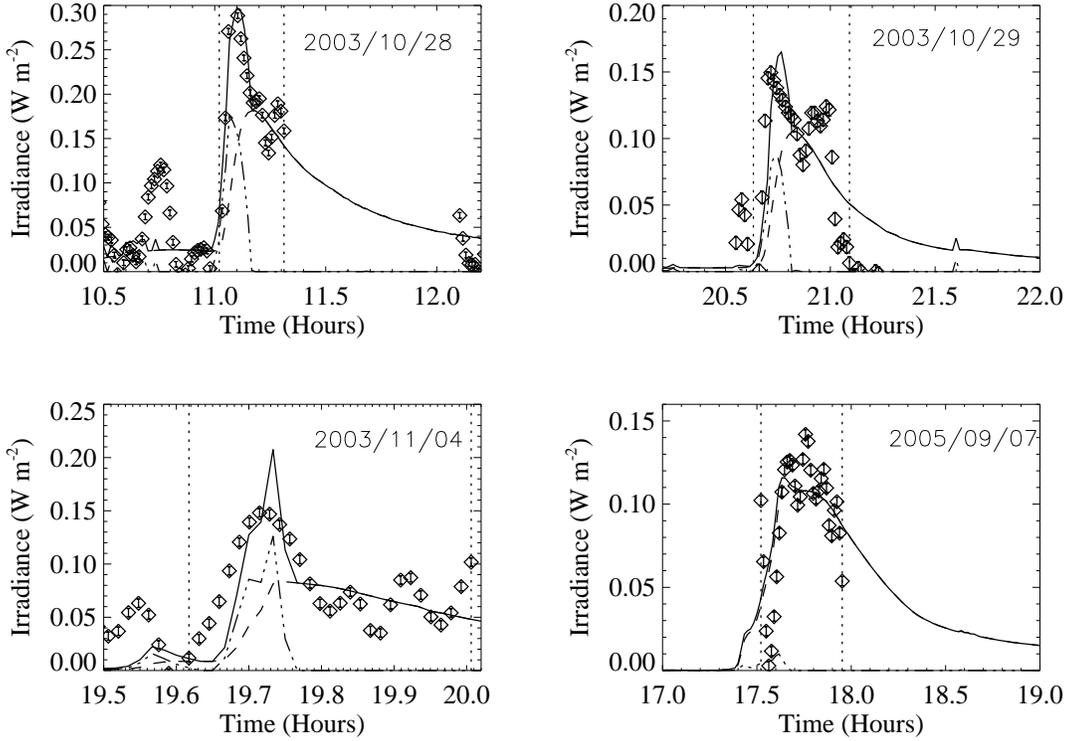}
              }
               \vspace{0.02\textwidth}   
\caption{
Background subtracted data for four of the five X class flares, X17 flare on 2003 October 28, X10 flare on 2003 October 29, X28 flare on 2003 November 4 and the X17 flare on 2005 September 7. The diamonds are the background subtracted TSI data with error bars comprised of instrument uncertainties added in quadrature with the linear background fit uncertainties. The dashed line is the scaled FISM gradual phase profile. The dashed-dot-dot-dot line is the scaled FISM impulsive phase profile. The solid line is the sum of the scaled impulsive and gradual phases. The vertical dotted lines are the timeframes of which the TSI data was fit to the FISM time profiles.}
\label{4_Flares}
\end{figure}


\clearpage



\begin{figure}      
   \centerline{\hspace*{-0.0\textwidth}
               \includegraphics[width=0.9\textwidth]{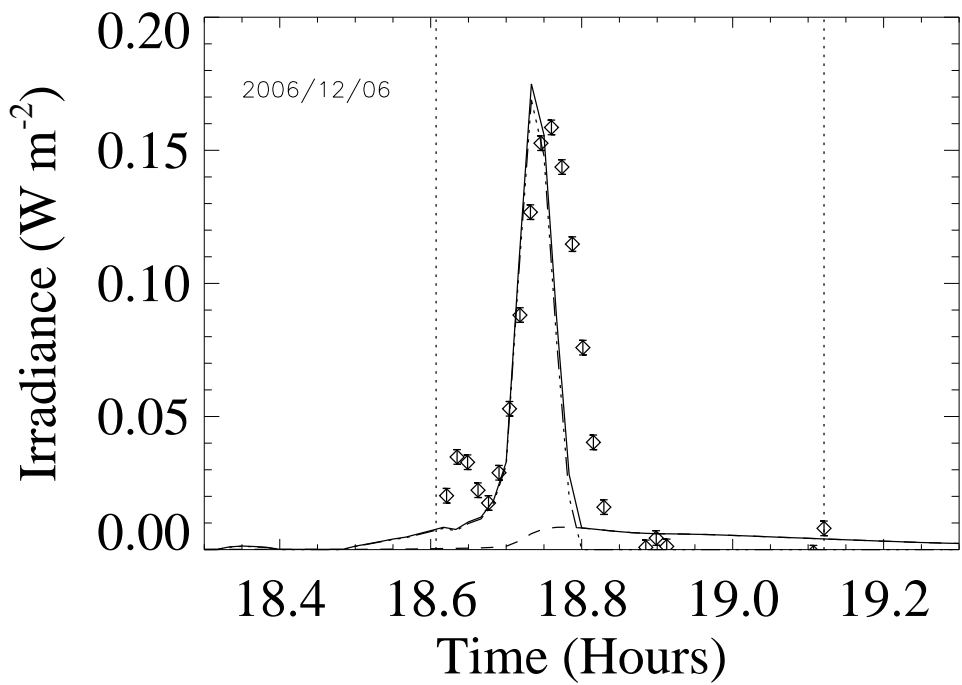}
              }
               \vspace{0.02\textwidth}   
\caption{
Background subtracted data for the X6.5 flare on 2006 December 6 not included in \cite{Woods2006}. The diamonds are the background subtracted TSI data with error bars comprised of instrument uncertainties added in quadrature with the linear background fit uncertainties. The dashed line is the scaled FISM gradual phase profile. The dashed-dot-dot-dot line is the scaled FISM impulsive phase profile. The solid line is the sum of the scaled impulsive and gradual phases. The vertical dotted lines are the time frames of which the TSI data was fit to the FISM time profiles.}
\label{Dec_6_2006}
\end{figure}


\clearpage



\begin{table}
\begin{center}
\caption{TIM Observed Flares (in $10^{30}$ ergs)\label{tbl-1}}
\resizebox{18cm}{!} {
\begin{tabular}{l*{10}{c}r}
\tableline\tableline
Date & Time & Class & Location & TSI-Tot & TSI-Imp & TSI-Grad &  VUV-Tot &  VUV-Imp & VUV-Grad \\
\tableline
2003/10/28 & 9:51 & X17  & E08 S16 & 626 & 87 & 539 & 112 & 7.0 & 106\\
2003/10/29 & 20:37 & X10 & W10 S17 & 346 & 44 & 302 & 49 & 5.2  & 43\\
2003/11/04 & 19:29 & X28  & W83 S19$^{a}$  & 290 & 32  &  258  & 89 & 2.8 & 86\\
2005/09/07 & 17.17 & X17  & E77 S11$^{a}$ & 436 & 6.2 & 430  & 76 & 2.6 & 74\\
2006/12/06 & 18:29 & X6.5 & E63 S06 & 76 & 53  & 23  & 20 & 2.1  & 18\\
\tableline
\end{tabular}
}
\tablecomments{The five X-Class flares with detected TIM TSI signatures from Solar Cycle 23. This list includes total bolometric energy release derived from the TSI time series fits. Also included are the VUV (1 - 190 nm) energy estimates obtained by integrating the FISM flare profiles. Both energy sets are decomposed into their impulsive and gradual phases.}
\tablenotetext{a}{Designates a limb flare (flare with an east/west position larger than 70).}
\end{center}
\end{table}


\clearpage



\begin{table}
\begin{center}
\caption{TIM Reconstructed Flares (in $10^{30}$ ergs)\label{tbl-2}}
\resizebox{18cm}{!} {
\begin{tabular}{l*{10}{c}r}
\tableline\tableline
Date & Time & Class & Location & TSI-Tot & TSI-Imp & TSI-Grad &  VUV-Tot & VUV-Imp & VUV-Grad\\
\tableline
2002/04/21 & 0:43 & X1.5  &W84 S14$^{a}$ & 112 & 3.8 & 108 & 26 & 1.0 & 25\\
2002/07/23 & 0:18 & X4.8 & E72 S13$^{a}$ & 102 & 7.2 & 95 & 24 & 1.9 & 22\\
2003/11/02 & 17:03 & X8.3  & E56 N14 & 148 & 43  &  105  & 47 & 4.2 & 43\\
2005/01/15 & 0:22 & X1.2  & E14 N8 & 8.4 & 2.1 & 6.3 & 2.8 & 0.2 & 2.6\\
2005/01/19 & 8:03 & X1.3 & W5 N15 & 64 & 16 & 49 & 22 & 1.5  & 20\\
2005/01/20 & 6:36 & X7.1 & W61 N14 & 218 & 40 & 178 & 77 & 3.9 & 73 \\
2006/12/05 & 10:18 & X 9.0 & E79 S07$^{a}$ & 166 & 19 & 147 & 39 & 5.0 & 34\\
2006/12/13 & 2:14 & X3.0 & W23 S05 & 113 & 37 & 76 & 35 & 3.6 & 31\\
\tableline
\end{tabular}
}
\tablecomments{The eight X-Class flares with reconstructed TSI values based on the FISM model flare profiles. The reconstructed TSI impulsive and gradual phases energies are obtained by scaling the VUV impulsive and gradual phase energies as described in Section 4.}
\tablenotetext{a}{Designates a limb flare (flare with an east/west position larger than 70).}
\end{center}
\end{table}




\end{document}